\documentclass{elphreport}
\usepackage{graphicx} % Include figure files

\begin{document}

\pacs{2633, 2638, 2658} %Put the ELPH experiment number, if you have.

\title{Report on strangeness photoproduction experiments performed with the Neutral Kaon Spectrometer 2. }

% \author[label1,label2]{}
% \address[label1]{}
% \address[label2]{}

\author[label1]{B. Beckford}
\author[label1]{A. Chiba} \author[label1]{D. Doi}  \author[label1]{J. Fujibayashi}\author[label1]{T. Fujii}\author[label1]{Y. Fujii} \author[label1]{K. Futatsukawa} \author[label1]{T. Gogami} \author[label1]{O. Hashimoto}

\author[label3]{Y.C. Han}

\author[label2]{K. Hirose} 

\author[label1]{S. Hirose} \author[label1]{R. Honda}  \author[label1]{R. Honda} \author[label1]{K. Hosomi}
\author[label1]{A. Iguchi}

\author[label2]{T. Ishikawa}

 \author[label1]{H. Kanda} \author[label1]{M. Kaneta} \author[label1]{Y. Kaneko} \author[label1]{S. Kato} 
 \author[label1]{D. Kawama} \author[label1]{T. Kawasaki}  \author[label1]{C. Kimura} \author[label1]{S. Kiyokawa} 
 \author[label1]{T. Koike} \author[label1]{K. Maeda}\author[label1]{K. Makabe} \author[label1]{N. Maruyama}
  \author[label1]{M. Matsubara} \author[label1]{K. Miwa} \author[label1]{Y. Miyagi} \author[label1]{S. Nagao} 
  \author[label1]{S. N. Nakamura} \author[label1]{A. Okuyama} \author[label1]{K. Shirotori} \author[label1]{K.Sugihara}
 
 \author[label2]{K.Suzuki}\author[label2]{T. Tamae} \author[label1]{H. Tamura} \author[label1]{N. Terada} \author[label1]{K. Tsukada} \author[label1]{K. Yagi}  
 \author[label1]{F. Yamamoto}  \author[label1]{T. O. Yamamoto}  
 
 \author[label2]{H. Yamazaki}
 
 \author[label1]{Y. Yonemoto}
%\author[label1,label2]{Third Author}
%\author[label1]{Fourth Author\thanksref{label3}}

\thanks[label3]{Present address: Research Center for Electron Photon Science, Tohoku
University, Sendai, 982-0826}

\address[label1]{Department of Physics, Tohoku University, Sendai, 980-8578}
\address[label2]{Research Center for Electron Photon Science, Tohoku University, Sendai,
982-0826}
\address[label3]{School f Nuclear Science and Technology, Lanzhou University, Lanzhou,
730000}

\maketitle

\begin{abstract}
An experiment designed to investigate the strangeness photoproduction process using a tagged photon beam in the energy range of 0.90 -1.08 GeV incident on a liquid deuterium target  was successfully performed. The purpose of the experiment was to measure the production of neutral kaons and the lambda particles on a deuteron.  The generation of photo produced particles was verified by the measurement of their decayed charged particles in the Neutral Kaon Spectrometer 2. The  reconstructed invariant mass distributions were achieved by selecting events where two or more particles tracks were identified. Preliminary results are presented here. 
\end{abstract}

\section{Introduction}
The reaction of interest discussed in this report  is the photoproduction of neutrally charged strange particles, ${\Lambda}$ and $K^{0}$, from a deuteron target.  The strangeness production processes by the electromagnetic interaction can be used  as a probe to bestow indispensable information on the strengths of meson-baryon coupling and internal structure of hadrons with the strangeness as a degree of freedom\cite{TMart} .  The exploration into the strangeness production mechanism in the threshold region allows for the frontier of physics to be explored particularly concerning the experimental measurement of the photo-produced $\Lambda$ particle's momentum and angular distribution.

As a result of the  series of the NKS experiments~\cite{Tsukada:2007}~\cite{Watanabe}, the re-envisioned  Neutral Kaon Spectrometer, NKS2, was newly designed and constructed at the Laboratory of Nuclear Science of Tohoku University (LNS) in 2004, replacing an original version, NKS. Its main purpose is to investigate the photo-production process, particularly the production of neutral strange particles via single $K^{0}$ and $\Lambda$ measurement with acceptance less biased at the forward region compared with that of the NKS spectrometer. It was further renovated in 2008 with the addition of a new inner detectors and hence forth is referred to as the NKS2+.

% Main text is here \cite{aaa}.
\section{The NKS2+ Spectrometer System}
The NKS2+  is composed of a cryogenic target system located in the center of various detector systems that works in tandem with a photon tagging array.  It is positioned along the fourth bending magnet (BM4) in the second experimental hall at the ELPH facility.  Using a carbon wire as a target, a photon beam is produced via the  electromagnetic interaction with the electrons orbiting in the Stretcher Booster Ring (STB) in the form of the bremsstrahlung radiation. The STB ring  is capable of accelerating electrons injected from a 0.15 GeV Linac injector up to 1.2 GeV. 
The photon beam is guided  through a collimator in order to reduce the beam halo, a sweep magnet, and into the NKS2+ and is bombarded on the target located in the spectrometer. Moving from the inner most position outwards, is the target which is surrounded by a Vertex Drift Chamber (VDC) and a  Inner Hodoscope comprised of twenty plastic scintillator segments (IH), which acts as the initiation time trigger for time of flight measurements. These pair of detectors are then surrounded and fully enclosed in a Cylindrical Drift Chamber (CDC).
  All detectors are themselves located in between the poles of a dipole magnet with 680 mm aperture.
 
  \begin{figure}[ht]
    \begin{center}
          \includegraphics[width=11.5cm]{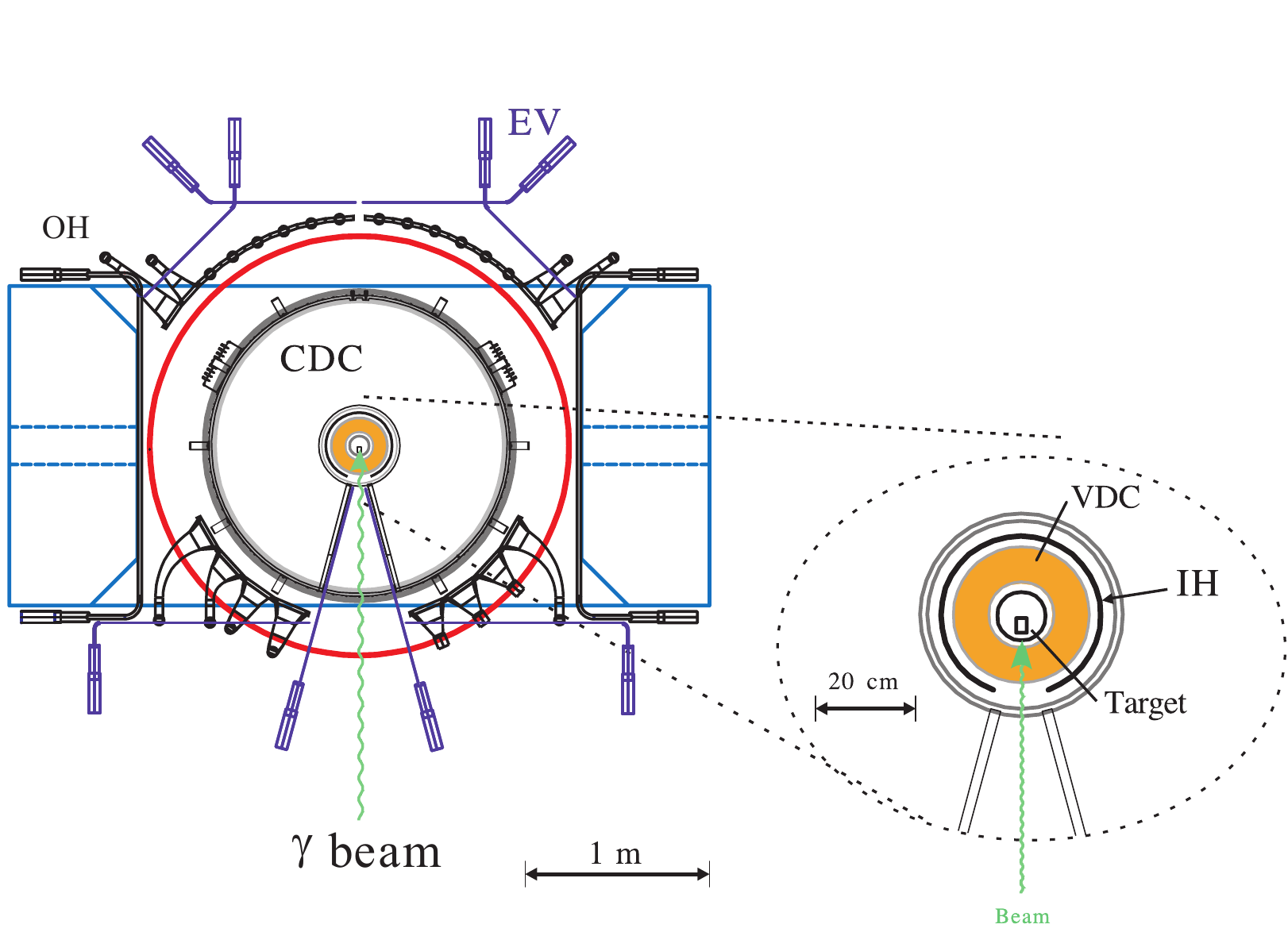}
      \caption{
        The figures shows a schematic perspective the recently improved NKS2 spectrometer. Here is a view  along the beam plane,where the $\hat{z}$ direction  = 0.  Inner detector package have been  redesigned and installed and is visible in the zoomed  image on the bottom right. 
      }
      \label{fig:NKS2_upgraded_detectors}
    \end{center}
  \end{figure}

\section{Data summary and analysis procedure}
\subsection{Trigger}
Our data acquisition utilized the  UNIDAQ on Linux as a core system . The main trigger utilized for physics production experiments was defined as equation~\ref{eq:trigger_logic}:
In the equation,  nTagSum is used to denote a  photon being generated and tagged with an over a range of 0.8$\le$$E_\gamma$$\leq$1.1 GeV, the energy of the photon is known by tagging the the recoil electron.
The trigger required that a minimum of a two particle event be detected in by the inner hodoscope(nIH) and the outer hodoscope(nOH). For the OH, a mixture of hits between the horizontal and vertical groups would be valid. Lastly, the EV is included into the trigger logic as a veto.
A summary of the data  analyzed for the results discussed in this report are listed in table~\ref{table:Production data}. 

  \begin{equation}
  \ {\rm Trigger} = ({\rm nTagSum}\geq1\otimes {\rm nIH}\geq2 )\otimes {\rm nOH}\geq 2\otimes \overline{\rm EV} 
          \label{eq:trigger_logic}
  \end{equation}

 \begin{table}
 \caption{Summary of data for physics production runs.}
\label{table:Production data} 
\begin{center}
\begin{tabular}{lccccc}
\hline
\hline

        Experiment Period     		& Flat Top  	 & Number of   			& Number of 			& Beam Rate \\
  	2010     			 	 & [s]    		& Accepted Events        	 & Photons [${\gamma}$]  	& [MHz]  \\
        \hline
        \hline
  %      July     				&  20    		& 0.16 x 10$^{9} $  	 	&  0.12  x 10$^{12}$    	&1.5 - 2.0 MHz\\
        September    			&  21  		& 0.64 x 10$^{9} $    		&  0.89 x 10$^{12}$		&1.5 - 2.5 MHz\\		
            
        \hline 
        \hline                                                                         
    %     TOTAL      				&          		& 0.64 x 10$^{9} $   		&  0.89 x 10$^{12}$	    	  &1.5 - 2.5 MHz \\
% \hline
% \hline
\end{tabular}
\end{center}
 \end{table}

\subsection{Time of flight}
Time of flight measurements (TOF)  allow for a particleÕs mass to be determined by knowledge of flight time, the momentum, and the path length L \cite{Nappi}. The time of flight for a particle exiting the inner hodoscopes and passing through the outer is found by the difference of the mean time of the outer with that of the inner. The time-of-flight (TOF) particle identification method in combination with a momentum measurement was used for determination of the particle species. The technique was to not select the particles only by their mass but instead by two dimensional histograms of the momentum versus inverse velocity (${\beta}$) was considered. From the energy-momentum relation the velocity can be calculated as,\\

 \begin{equation}
  E^2 = m^2 + p^2
 \end{equation}
 
\begin{equation}
{\beta}  = \frac{p}{E} = \frac{p}{\sqrt{m^{2}+p^{2}}}
\end{equation}
Therefore the square of the particle's mass is found by,

\begin{equation}
\label{eq:mass_squared}
m^2 = (\frac{p}{{\beta}})^2 = p^2(\frac{1}{{\beta^2}-1})
      \label{eq:mass_calculation}
\end{equation}

\begin{figure}[htbp]
\begin{center}
	      \includegraphics[width=12.5cm]{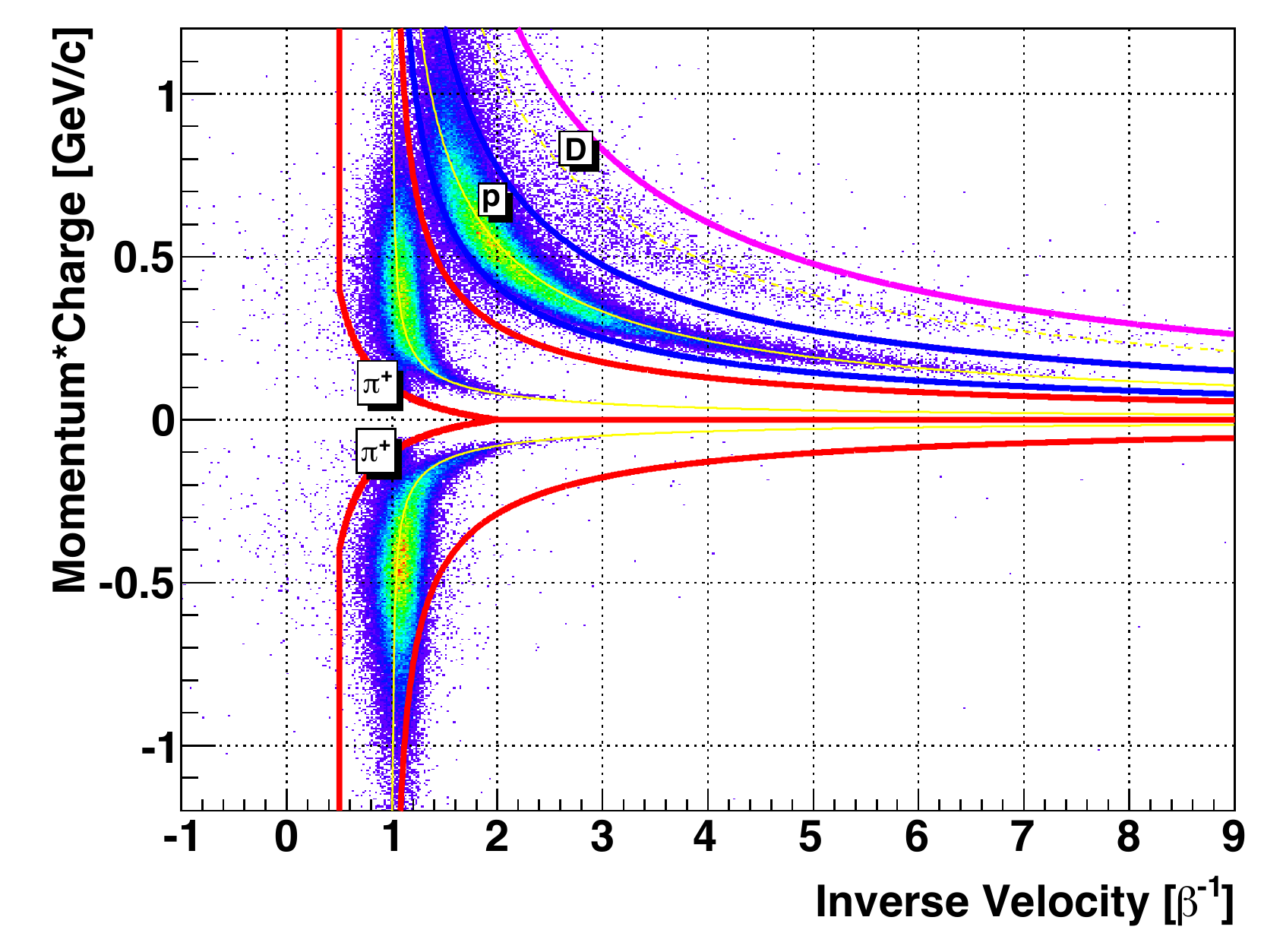}
     	 	\caption{The particle momentum multiplied with the particle charge plotted as a function of the inverse velocity is shown in figure.~\ref{fig:momentum_versus_inverse_beta}.  The proton and pion selection regions are shown by the red and blue  regions respectively. The charged pions are identified by the sign of the momentum.  }
     	 \label{fig:momentum_versus_inverse_beta}
\end{center}
\end{figure}

Figure.~\ref{fig:momentum_versus_inverse_beta} displays the two dimensional plot of the charge of the particle multiplied by momentum versus inverse beta. The colored locus plot was produced  with an additional limitation on the opening angle between particles; The requirement of  -0.9 $\le $cos${\theta}$ $\le$ 0.9 was applied for the purpose of rejecting the abundant amount of electron positron pairs. The bands around 1/${\beta}$ = 1 are the distributions that correspond to charged pions, indicated in the red lined regions,  and the band is for protons which lie within the blue lined region. The time of flight resolution between the vertical segments of the OH and IH was found to be 326 (${\sigma}$)$\pm$ 2 ps. For the  the left and right second segments of the inner hodoscopes (IH2), which are used in the trigger logic for data acquisition, the relative time resolution was 170 (${\sigma}$) $\pm$ 1 ps.

\begin{figure}
\begin{center}
         \includegraphics[width=12cm]{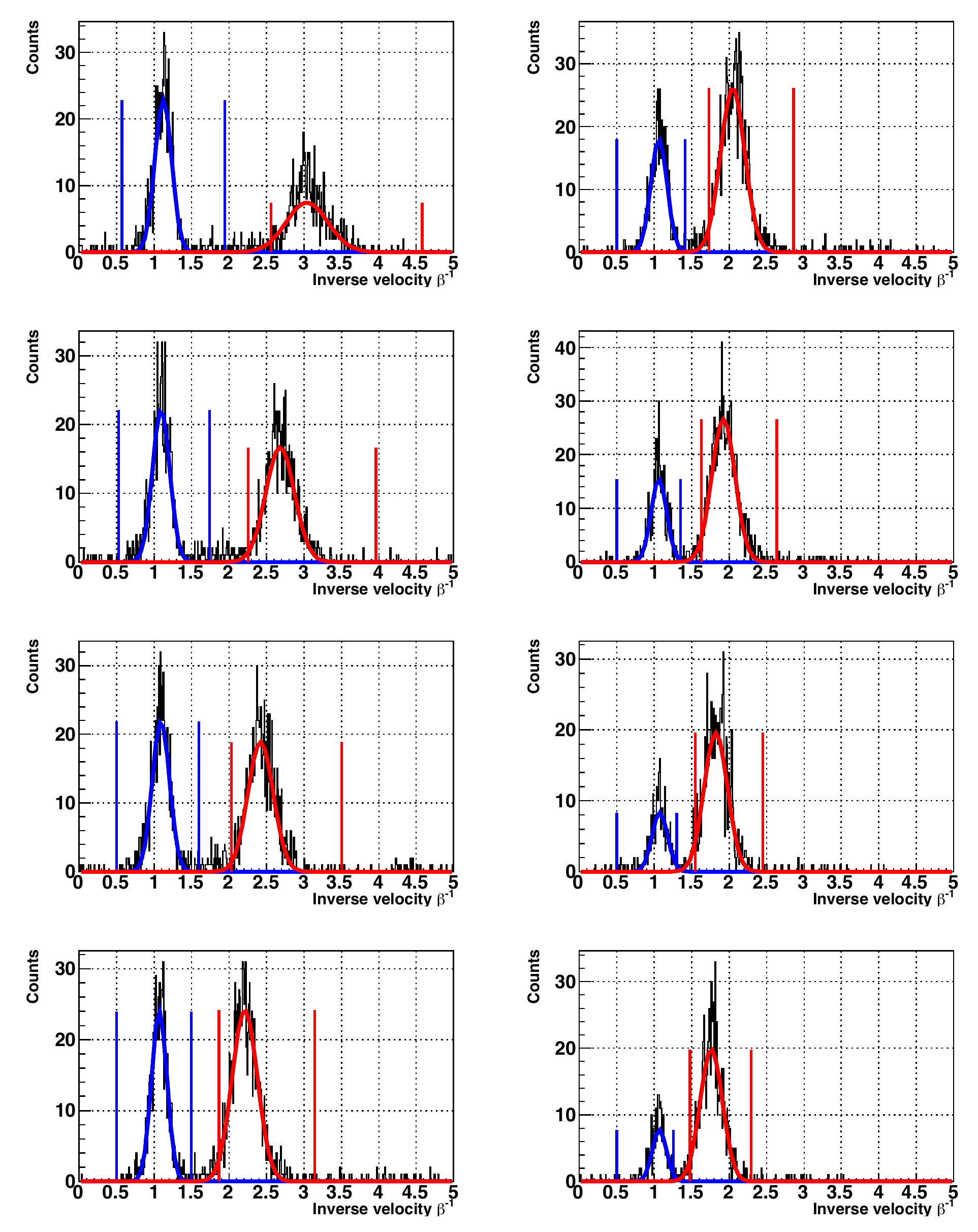} 
    \caption{ 1/${\beta}$ distributions, where ${\beta}$ = (v/$c$).
      The figures are for a momentum selection region of 0.02-0.22, 0.3-0.32, 0.4-0.4.2,0.5-0.52 ,0.60-0.62, and 0.7-0.72 GeV/$c$  
      }
  	  \label{fig:beta_distributions} 
\end{center}
\end{figure}

\begin{figure}
\begin{center}
	\includegraphics[width=11cm]{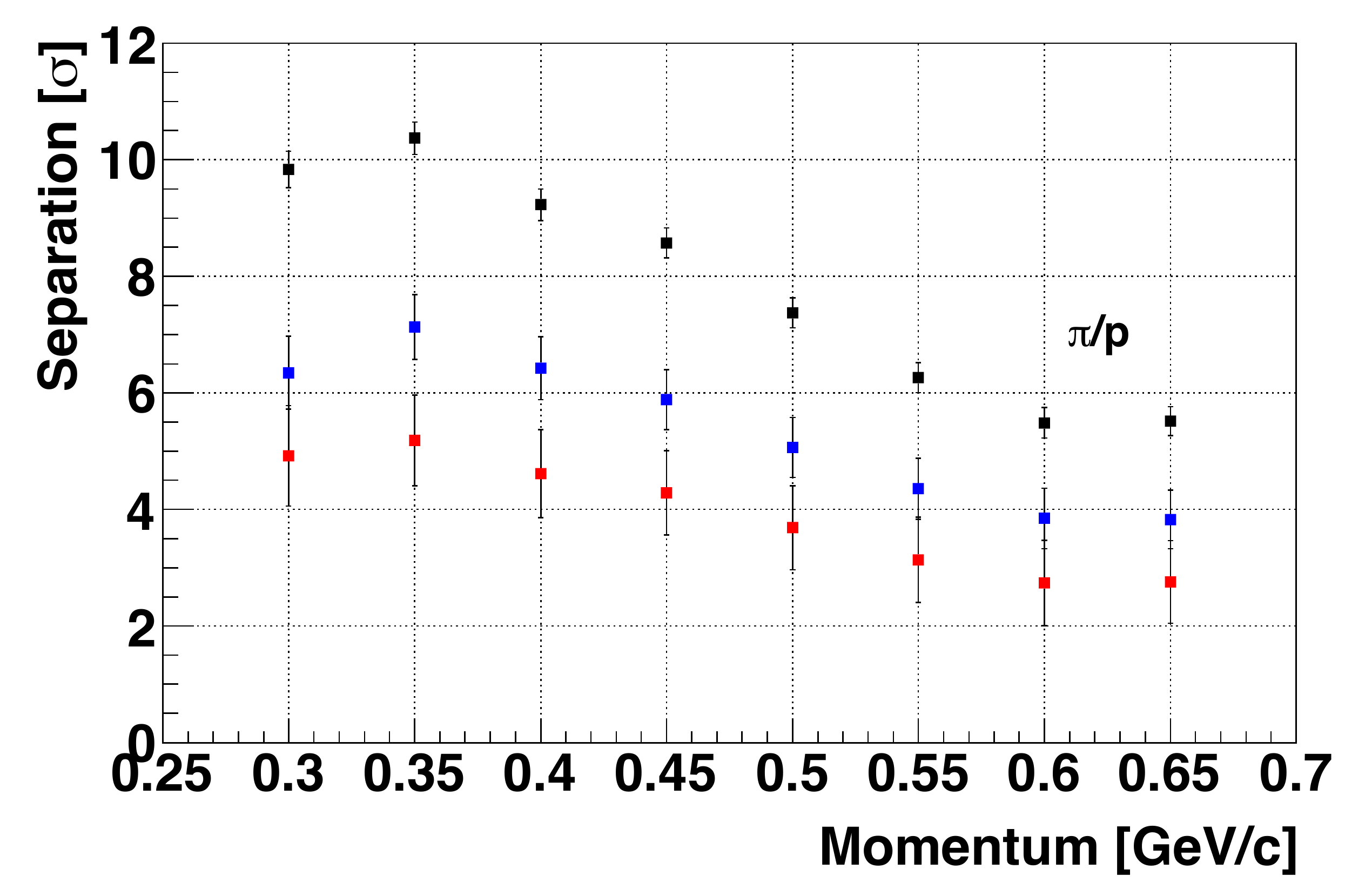}
	\caption{Separation power between pions and protons with respect to momentum for momentum selected regions of 0.02-0.25, 0.3-0.35, 0.4-0.4.5,0.5-0.55 ,0.60-0.65, and  0.7-0.75 GeV/$c$. The results of equations \ref{eq:sep_pow1}, \ref{eq:sep_pow2} , \ref{eq:sep_pow3} are shown as the black , blue and red data points respectively. }
	\label{fig:separation_power}
\end{center}
\end{figure}

A study into the separation power of time of flight measurements between ${\pi}$and protons was conducted to quantitatively ascertain the performance of the NKS2 spectrometer with the inclusion of the new inner hodoscope (IH).
The separation power of the particle species is evident when a projection of the inverse velocity distribution is produced for a specified momentum region as presented  in figure.~\ref{fig:beta_distributions}, where the pion and proton inverse beta distributions are seen left to right respectively.  The regions of momenta selection are 0.02-0.25, 0.3-0.35, 0.4-0.4.5,0.5-0.55 ,0.60-0.65, and  0.7-0.75 GeV/$c$.  This plot includes an additional requirement for the opening angle such that $-0.9 \leq \cos\theta \leq 0.9$. The opening angle for electron positron pairs is much smaller than that of charged pions from the kaon decay in the energy region of this experiment. The results of the study of the separation power between pions and protons with respect to momentum was performed and the extracted results are shown in figure~\ref{fig:separation_power}. The separation power $n_{{\sigma_i}}$ was defined by  the following equations \ref{eq:sep_pow1}, \ref{eq:sep_pow2} , \ref{eq:sep_pow3}, where the distributions in figure~\ref{fig:beta_distributions} were fitted to a Gaussian function. $\mu_1$, $\mu_2$,  $\sigma_1$ and  $\sigma_2$ are the mean values  and the standard deviations of the gaussian fits of the pion and proton distributions. The calculation of  the separation power $n_{{\sigma_i}}$ is,

\begin{equation}
n_{{\sigma_1}}=\frac{|{\mu_1+\mu_2}|}{({\sigma_1+\sigma_2})/2}
\label{eq:sep_pow1}
\end{equation}
\begin{equation}
n_{{\sigma_2}}=\frac{|{\mu_1+\mu_2}|}{\sqrt({{\sigma_1}^{2}+{\sigma_2}}^{2})}
\label{eq:sep_pow2}
\end{equation}
\begin{equation}
n_{{\sigma_3}}=\frac{|{\mu_1+\mu_2}|}{({\sigma_1+\sigma_2})}
\label{eq:sep_pow3}
\end{equation}

As can be understood from the equations the particles species separation power was defined as being proportional to the mean of the sigmas, equation \ref{eq:sep_pow1}, the square root of the sum of the squares of the sigmas, equation \ref{eq:sep_pow2}, and  the sum of the sigmas, equation\ref{eq:sep_pow3}. Therefore, we define the potential upper bound and lower bound separation between protons and pions as equations \ref{eq:sep_pow1} and ~\ref{eq:sep_pow3}. 

From equation ~\ref{eq:mass_squared} the region selection of the identification of  charged pions is specified as,

\begin{equation}
-0.5 \le m^2 \le 0.25 (GeV^2/c^4)
\end{equation}

and for the protons,

\begin{equation}
0.5 \le m^2 \le1.8 (GeV^2/c^4)
\end{equation}

\begin{figure}
\begin{center}
		 \includegraphics[width=7.cm]{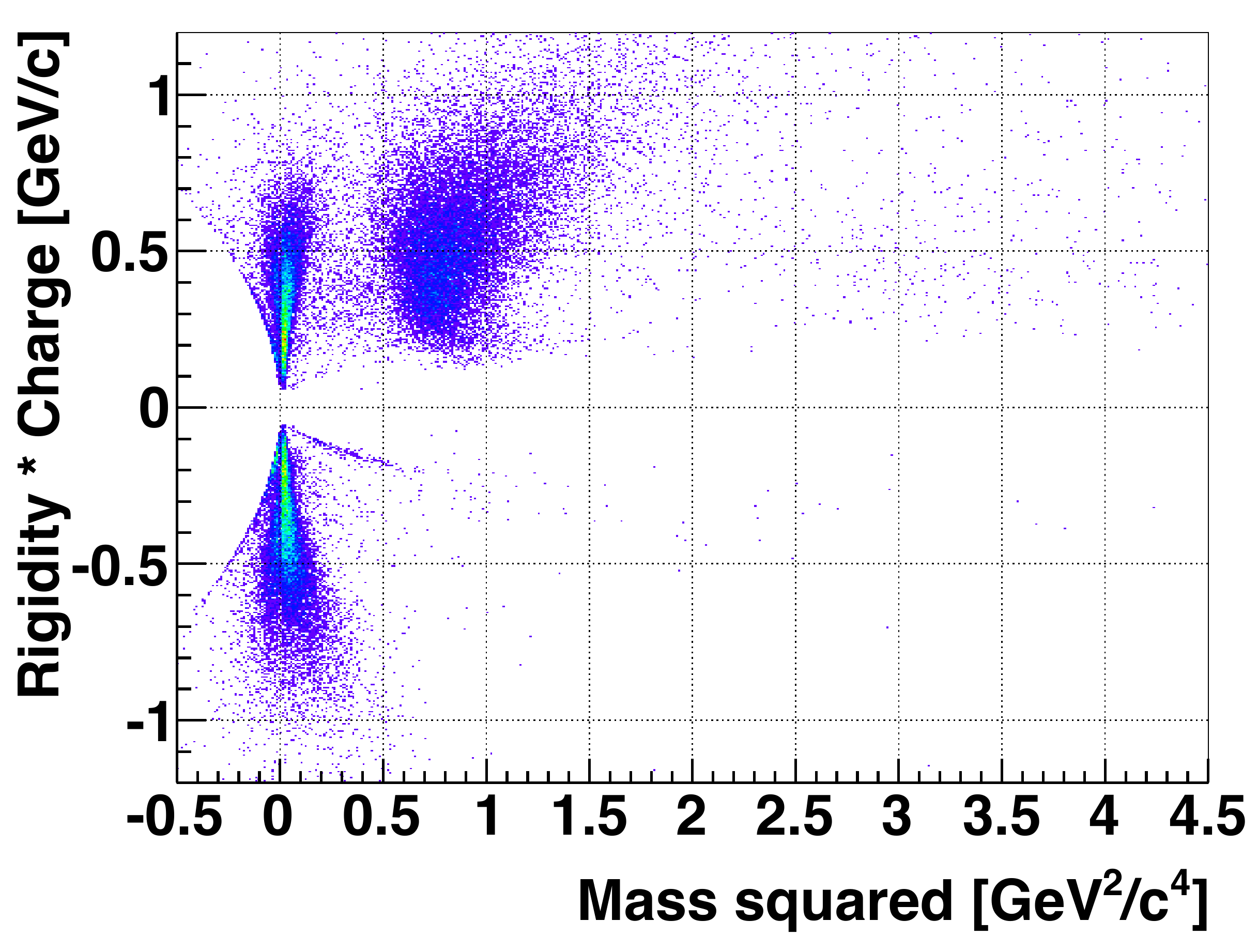}
    	     	\includegraphics[width=7.cm]{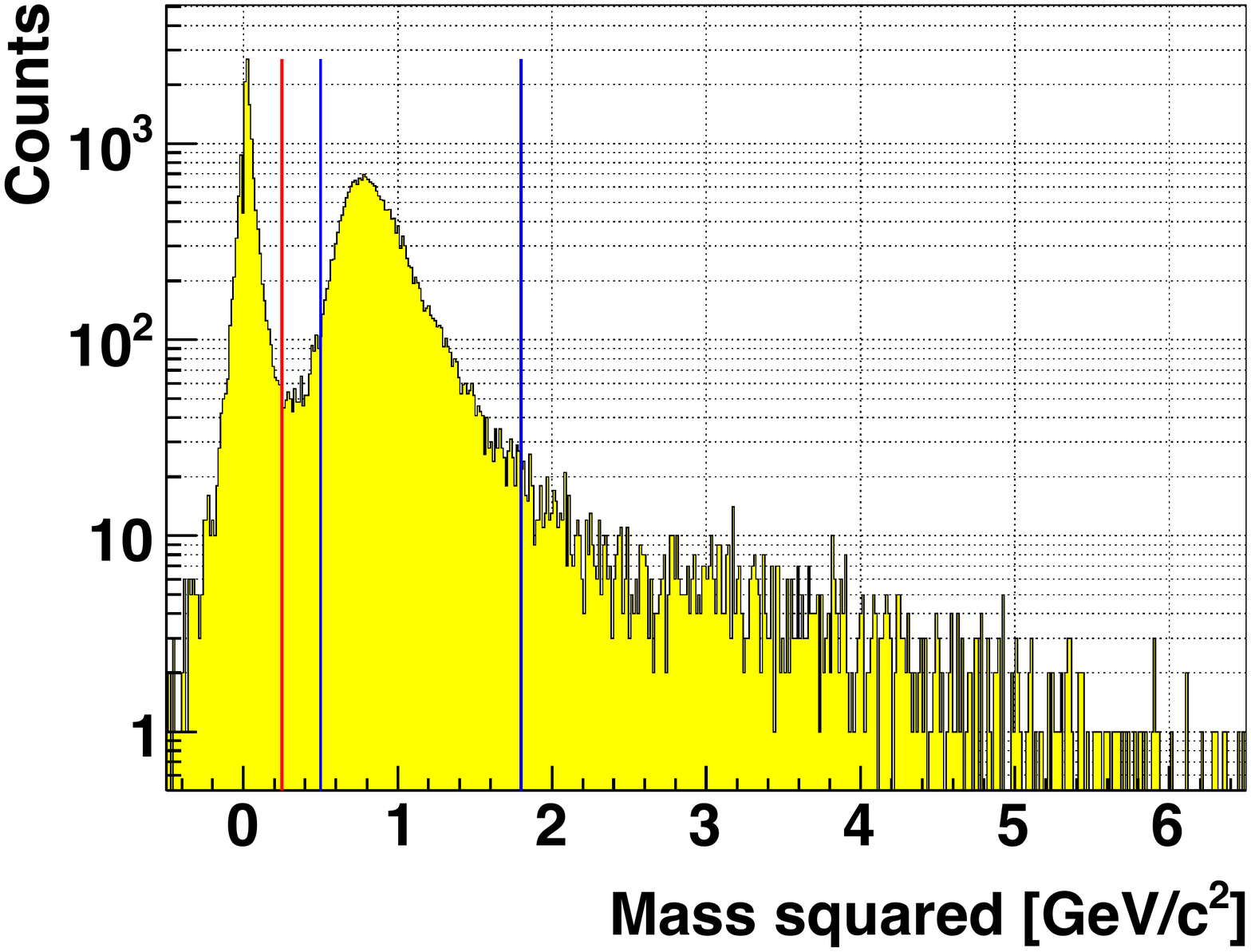}  
	         \caption{Rigiidity multiplied by particle charge plotted verses mass squared event distribution(left); Mass squared  event distribution (right). }
      		\label{fig:momentum_versus_mass_squared}
\end{center}
\end{figure}

The mass squared distributions of the detected particles  versus the momentum that has been multiplied by the charge is shown in figure~\ref{fig:momentum_versus_mass_squared}.

\section{Invariant Mass}
For the identification of the lambda particle it was necessary to reconstruct its invariant mass from two particle combinations of a negatively charged pion and a proton. The invariant mass is calculated by the following equation,
 \begin{equation}
  	\label{eq:ppi_invariant_mass_squared}
 	M_{{p}{\pi^{-}}^{2}} = ( \sqrt{m_{p}^2+|\vec{p}_{p}|^2 }+ \sqrt{m_{\pi^{-}}^2+|\vec{p}_{\pi^{-}}|^2 } )^2 - |\vec{p}_{p} + \vec{p}_{\pi^{-}}|^2
 \end{equation}

\begin{figure}[htbp]
\begin{center}
\includegraphics[width=8cm]{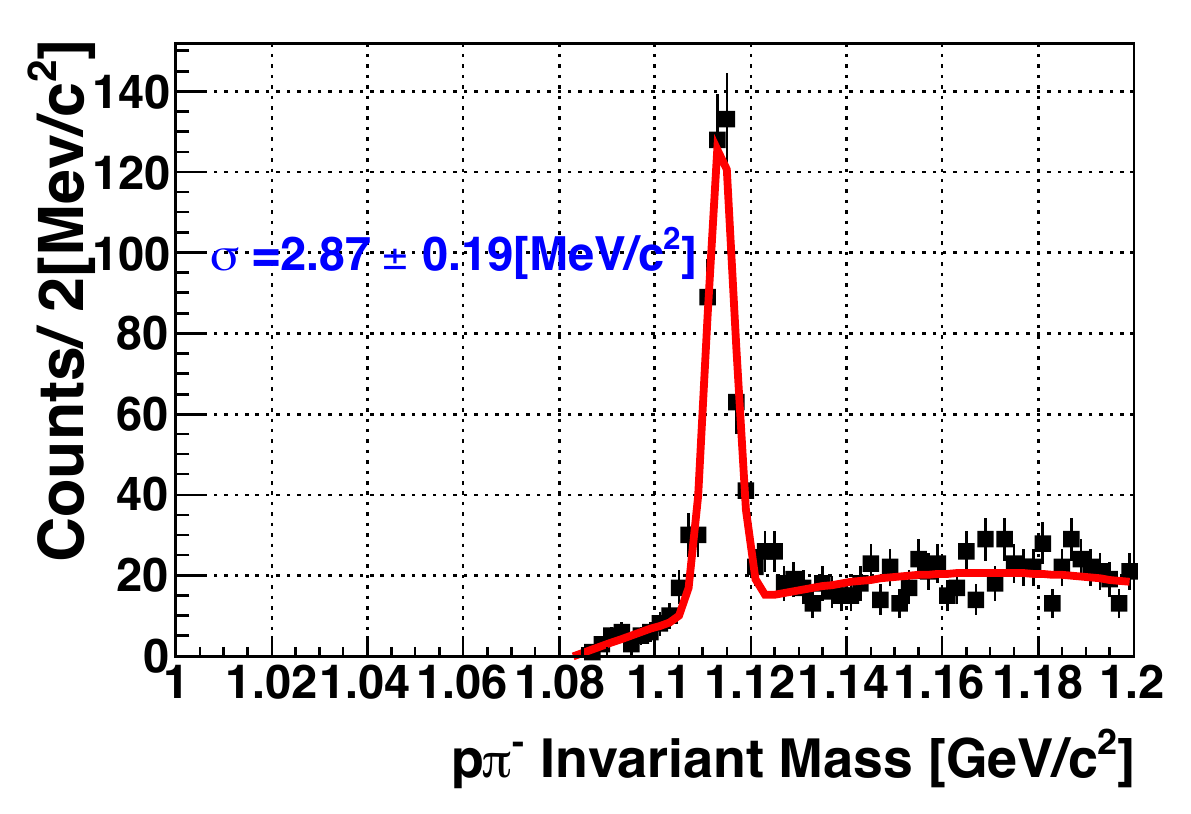}
\caption{Invariant mass resolution in rms for the energy bin of 0.9 -1.1 GeV was found by fitting as 2.87 $\pm$ 0.19 MeV/c}
\label{fig:im_fitted_E_all}
\end{center}
\end{figure}

\begin{figure}[htbp]
\begin{center}
\includegraphics[width=14cm]{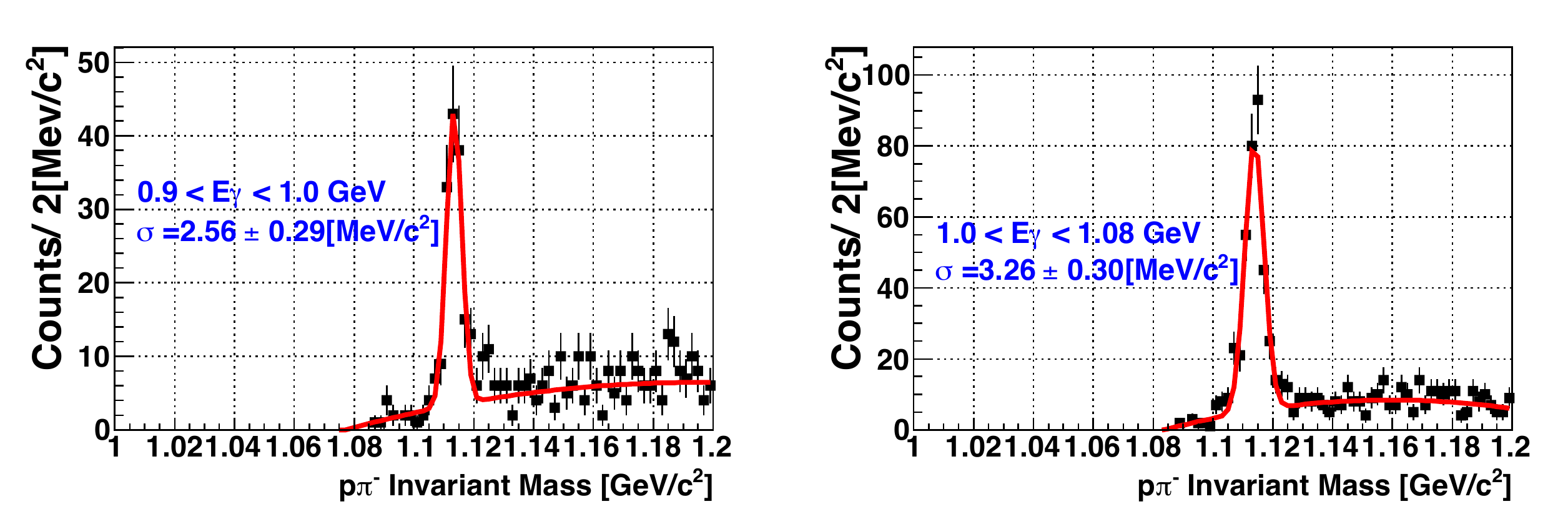}
\caption{Invariant mass resolution in rms for the energy bins of 0.9-1.0 and 1.0 -1.1 GeV are seen in the figures left to right respectively., where the fitted resolution were found to be 2.56 $\pm$ 0.29 MeV/c and 3.26 $\pm$ 0.30 MeV/c }
\label{fig:im_fitted_E_bins}
\end{center}
\end{figure}

 The p${\pi-}$invariant mass distribution in the inclusive ${\gamma}$N $\rightarrow$p${\pi}^{-}$X measurement  is shown figure~\ref{fig:im_fitted_E_all} for the energy bin of 0.9 -1.08 GeV. The distributions associated with the energy bins of 0.9-1.0 and 1.0 - 1.08 GeV are can be seen in figure~\ref{fig:im_fitted_E_all}. However, there lies the chance of a fallaciously identified decay vertex point such that the combination of a pion from the kaon decay and a proton from a lambda decay. The background present in the invariant mass spectrum after the application of kinematical and various cuts is attributed mainly to this mis-combination. The photon induced production of a ${\Delta^{0}}$ particle and it's subsequent decay can also contribute to the background. 
The distributions were fitted by the combination of a gaussian and third degree polynomial function. The resolution of the distribution generated for the energy bin  of 0.9 -1.08 GeV was determined to be 2.87 $\pm$ 0.19 MeV/c$^2$ in RMS. In the two smaller energy bins, 0.9-1.0 and 1.0-1.1 GeV,  the resolutions were  2.56 $\pm$ 0.2 and 3.26 $\pm$ 0.30 GeV respectively. Full details are listed in table~\ref{table:lambda_invariant_mass_resolution}.
  \begin{table}
 \caption{p${\pi}^{-}$ invariant mass resolution }
 \label{table:lambda_invariant_mass_resolution}
 \begin{center}
 \begin{tabular}{cccccc}
 \hline
 \hline

 							&										&Inclusive ${\Lambda}$ measurement \\
    \hline
    \hline							
Selection						& Region	[ GeV]								& Resolution	[MeV/c$^2$]	&Yield (S)			&Noise(N) 			&S/N	\\
 \hline
E$_{\gamma}^{all}$ 				& 0.90 $\le$ E$_{\gamma}$ $\le$1.1  			& 	2.87 $\pm$ 0.19			&349 $\pm$ 25.49	&166$\pm$12.88		&2.1\\				
E$_{\gamma}^{1}$ 				& 0.90 $\le$ E$_{\gamma}$ $\le$1.0  			& 	2.56 $\pm$ 0.20			&102 $\pm$ 14.42	&49$\pm$7.00			&2.1\\	
E$_{\gamma}^{2}$ 				& 1.0  $\le$ E$_{\gamma}$ $\le$1.1   			& 	3.26 $\pm$ 0.30			&247 $\pm$ 20.59	&72$\pm$10.24			&3.4\\
 \hline
 \hline		
 		
 \end{tabular}
 \end{center}
 \end{table}
		 
% Subsection \cite{aaa,bbb}.

%\subsubsection{Subsubsection}
 %Figure \ref{FIG:fig} is here.
 %\begin{figure}[!ht]
  %\begin{center}
   %\includegraphics[width=0.3\textwidth]{fig1.eps}
   %\caption{This is a caption}
   %\label{FIG:fig}
  %\end{center}
 %\end{figure}

 \section{Summary }
 
 The experiment in the photoproduction of strangeness on a deuteron target was accomplished using a tagged photon beam at the ELPH research facility.
The time of flight particle identification (PID) capability of the upgraded NKS2+ spectrometer was evaluated. The separation power between protons and pions was examined and reported.
 The invariant mass spectrum of p${\pi}^-$ was obtained and the resolution of the spectrums have been detailed.
We are placing immense effort in the proper calibration of the data and in furthering the status of data analysis, which  is still underway.

\section*{Acknowledgment}
This research endeavor was partially supported by the Creative Research Program funded by the Japan Promotion of Science and also by the JSPS Core to Core program.

\end{document}